\documentclass[prb,aps,showpacs,twocolumn,epsf,nobinotes]{revtex4}
\usepackage{graphicx}% Include figure files
\usepackage{bm}% bold math
\usepackage{amsfonts,amsmath,amssymb}
\usepackage[squaren]{SIunits}
\makeatletter
\begin{document}
\title{Magnetotransport properties in purple bronze
Li$_{0.9}$Mo$_6$O$_{17}$ Single Crystal}
\author{H. Chen, J. J. Ying, Y. L. Xie, G. Wu, T. Wu}
\author{ X. H. Chen }
\altaffiliation{Corresponding author} \email{chenxh@ustc.edu.cn}
\affiliation{Hefei National Laboratory for Physical Science at
Microscale and Department of Physics, University of Science and
Technology of China, Hefei, Anhui 230026, People's Republic of
China}

\begin{abstract}
We have measured resistivity along the a, b and c axes of
Li$_{0.9}$Mo$_6$O$_{17}$ single crystal. The anisotropy
$\rho$$_c$/$\rho$$_a$ and $\rho$$_c$/$\rho$$_b$ is given, confirming
the quasi-one-dimensionality of the compound. The sharp decrease in
the anisotropy below a certain temperature (T$_M$) indicates
dimensional crossover. Superconductivity occurs at 1.8 K well below
$T_M$. Negative MRs are observed with magnetic field (H) applied
along b axis. This could be ascribed to suppression of energy gap
associated with CDW state. While large positive MR is observed with
H $\parallel$ c-axis. The MR data can be well fitted by a modified
two-band model which has been used in CDW compounds such as
quasi-two-dimensional purple bronzes A$_{0.9}$Mo$_6$O$_{17}$ (A = K,
Na, Tl) and quasi-one-dimensional conductor NbSe$_3$. The behavior
of MR provides a strong evidence for the existence of CDW
instability in $Li_{0.9}Mo_6O_{17}$.
\end{abstract}
 \maketitle

\section{INTRODUCTION}

The Fermi liquid theory is successful in describing the behavior of
the ordinary metals, but breaks down in attempts to explain
properties of one-dimensional (1D) systems. In a 1D interacting
electron system, the low-energy physics is described by the
Luttinger liquid (LL) picture.\cite{Luttinger, J.Voit} Two key
features of the Luttinger liquid are the spin-charge separation,
where the spin and the charge degrees of freedom are completely
separated into collective density waves (¡°spinons¡± and ¡°holons¡±)
that propagate with different velocities, and in approaching the
Fermi energy $E_F$, the energy dependence of the momentum-summed
single-particle density of states (DOS) displays a power-law decay
with an anomalous exponent, i.e.,
$\mid$E$\mid$$^\alpha$.\cite{J.Voit} However, an arbitrarily small
perturbation will destabilize the Luttinger liquid fixed point. when
$0 < \alpha < 1$, $Li_{0.9}Mo_6O_{17}$ is in this case, instability
will lead to higher-dimensional behavior. Below the crossover
temperature $T_{1D}$ $\sim$ ($t_\perp$)$^{1/(1-\alpha)}$, the LL
fixed point may be destabilized via single-fermion hopping (for
smaller $\alpha$) or two-fermion hopping (for larger $\alpha$). The
former leads to a Fermi liquid, while the latter leads to
semiconductor charge density wave (CDW) or spin density wave (SDW)
behavior provided that the electron-electron interaction is
repulsive. When it is attractive, singlet or triplet superconducting
states show up.\cite{Daniel. Boies,dos Santos}

The so-called purple bronze $Li_{0.9}Mo_6O_{17}$ has been studied by
many scientists for more than two decades. Other members of purple
bronze A$_{0.9}Mo_6O_{17}$ (A = K, Na, Tl ) have a 2D crystal
structure giving rise to 2D metallic properties.\cite{Buder,
Vincent, Greenblatt, Ganne, Ramanujachary, McCarroll} While
$Li_{0.9}Mo_6O_{17}$, which has a 3D crystal structure, has been
confirmed a quasi one-dimensional (1D) system having highly
anisotropic electronic properties.\cite{M. Greenblatt,Choi} It does
not contain separated layers of composition
Mo$_6$O$_{17}$.\cite{Whangbo}

\begin{figure}[h]
\includegraphics[width=9.5 cm]{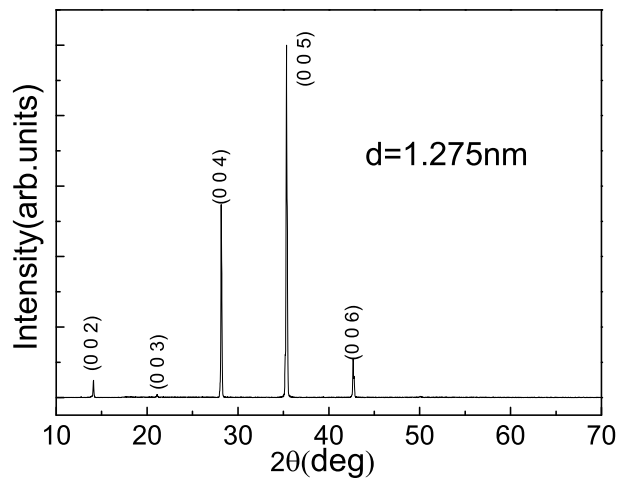}
\caption{Single crystal x-ray diffraction pattern of
$Li_{0.9}Mo_6O_{17}$}\label{Fig:Fig1}
\end{figure}

In $Li_{0.9}Mo_6O_{17}$, a superconducting transition occurs at
$\sim$ 1.8 K and a metal-insulator transition ( MIT ) occurs at
T$_M$ $\sim$ 30 K.\cite{Greenblatt1,Schlenker} The cause of the MIT
near T$_M$ $\sim$ 30 K is an open question. Considering that other
members of purple bronze A$_{0.9}$Mo$_6$O$_{17}$ ( A = K, Na, Tl )
are well known electron-phonon coupling induced CDW
compounds,\cite{C. Schlenker} the MIT in $Li_{0.9}Mo_6O_{17}$ at
T$_M$ has been attributed to CDW (or SDW) instability. But there is
no evidence for the SDW/CDW either in magnetic susceptibility or
optical experiments.\cite{Choi} Furthermore there are contradictory
reports from angle resolved photoemission spectroscopy (ARPES) on
Li$_{0.9}$Mo$_6$O$_{17}$: some support LL behavior,\cite{Denlinger,F
W} while Xue et al. reported that above T$_M$ a peak dispersing to
define its Fermi surface develops Fermi energy (E$_F$) weight
requiring a Fermi liquid (FL) description, and below T$_M$ a large
gap was observed, being consistent with CDW behavior.\cite{Jinyu
Xue} Localization was also put forward as possible origin of the
upturn in resistivity below T$_M$.\cite{Choi} But if the upturn at
T$_M$ is from localization, it would probably preclude
low-temperature superconductivity due to the strong disorder
effects. Santos et al. considered that T$_M$ represents a crossover
in dimensionality. They studied the thermal expansion of three axis
on single crystal $Li_{0.9}Mo_6O_{17}$, and concluded that the
upturn of resistivity response in $Li_{0.9}Mo_6O_{17}$ is due to a
CDW instability dominated by electronic interactions. The
dimensional crossover at $T_M$ may destabilize the LL fixed point,
and leads to a CDW dominated by electronic interactions as suggested
by LL theories.\cite{dos Santos,Daniel. Boies} Anyway, the
understanding of the MIT at $\sim$ 30 K have reached no consensus up
to now.

\begin{figure}[h]
\includegraphics[width=9.5 cm]{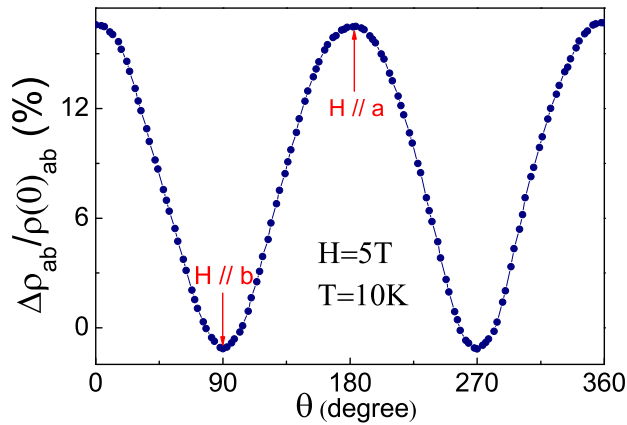}
\caption{Angular dependence of ab-plane MR at T = 10 K and H = 5 T
with twofold symmetry. The field was rotated within the ab-plane.
}\label{Fig:Fig1}
\end{figure}

In this paper, we have studied resistivity along the a, b and c axes
on the same piece of $Li_{0.9}Mo_6O_{17}$ single crystal,
respectively. The anisotropy of $\rho$$_c$/$\rho$$_a$ and
$\rho$$_c$/$\rho$$_b$ was given, confirming the
quasi-one-dimensionality of the compound. The sharp decrease in the
anisotropy below T$_M$ indicates a dimensional crossover at $T_M$.
It is striking that a negative MR is observed with magnetic field
applied along b axis, while huge positive MR shows up with H
$\parallel$ c. The MR data can be well fitted by a modified two-band
model which had been widely used in CDW quasi-two-dimensional purple
bronzes A$_{0.9}$Mo$_6$O$_{17}$ ( A = K, Na, Tl ) and
quasi-one-dimensional conductor NbSe$_3$.\cite{Mingliang Tian, J. Q.
Shen} It gives evidence for the existence of CDW instability in
$Li_{0.9}Mo_6O_{17}$.

\section{EXPERIMENTAL PROCEDURE}

\begin{figure}[h]
\includegraphics[width=9.5 cm]{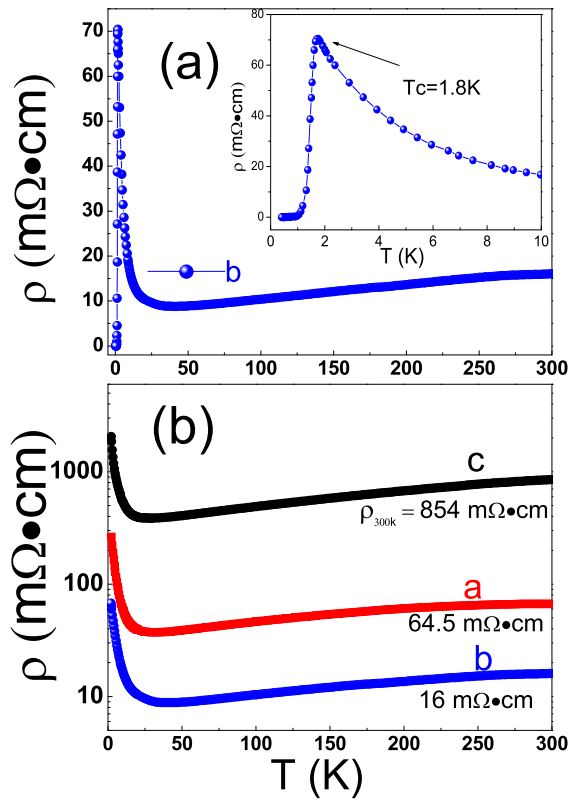}
\caption{(a): Temperature dependence of electrical resistivity
$\rho$ in the temperature range from 0.3 K to 300 K and the inset
highlights the superconducting transition in the temperature range
from 0.3 K to 10 K for $Li_{0.9}Mo_6O_{17}$ single crystal along b
axis. (b): Temperature dependence of electrical resistivity for the
three crystallographic directions for the same piece of
$Li_{0.9}Mo_6O_{17}$ single crystal. The $\rho$$_{300K}$ values for
each axis are shown}\label{Fig:Fig1}
\end{figure}

The $Li_{0.9}Mo_6O_{17}$ single crystals were grown by
temperature-gradient flux method\cite{McCarroll, dos Santos} using
Li$_2$MoO$_4$, MoO$_2$ and MoO$_3$ as starting materials.
Li$_2$MoO$_4$ was presynthesized by heating Li$_2$CO$_3$ and MoO$_3$
in stoichiometric ratio at 500$\celsius$, 600$\celsius$, and
650$\celsius$ in air for 3 days with several intermediate grinds.
MoO$_2$ was presynthesized by heating Mo and MoO$_3$ in an evacuated
quartz tube at 800$\celsius$ for 36 hours. These products were
weighted according to the raw proportion as in reference
[10]\cite{McCarroll} and thoroughly ground in an argon-filled glove
box, then sealed in an evacuated 15 cm long quartz tube, and held at
575$\celsius$ for 4 days, then reacted 10 days in a gradient of
10$\celsius$/cm (490$\celsius$ and 640$\celsius$ at the tube ends)
and finally cooled to room temperature by furnace. The mixture was
immersed in hot 5$\%$ potassium carbonate until the solutions
obtained were colorless. Typical dimension of the crystals is
2$\times$1$\times$0.25 mm$^3$ with surface color of purple and
bronze.

Single crystals were characterized by x-ray diffractions (XRD) using
Cu $K_{\alpha}$ radiations. As shown in Fig.1, only (00l)
diffraction peaks were observed, it suggests that the
crystallographic c-axis is perpendicular to the plane of the
plate-like single crystal. The lattice constant of c-axis was
determined to be 1.275 nm.

The electrical transport properties were measured using the ac
four-probe method with an alternative current (ac) resistance bridge
system (Linear Research, Inc.; LR-700P) in the range 0.3 K $<$ T $<$
300 K along the a, b, c axis, respectively. The dc magnetic field
for MR measurements is supplied by a superconducting magnet system
(Oxford Instruments). Crystal orientation was determined through
angular dependence of ab-plane magnetoresistance(AMR). Magnetic
field was rotated within ab-plane. It is found that the AMR shows a
twofold symmetry. As shown in fig.2, we define the direction with
the maximum MR as a-axis and the direction with minimum MR as
b-axis.

\section{EXPERIMENTAL RESULTS AND DISCUSSION}

Fig.3 (a) displays temperature dependence of resistivity for single
crystal $Li_{0.9}Mo_6O_{17}$ along b axis. As shown in Fig.3(a), a
metal-insulator transition is observed at T$_M$ $\sim$ 30 K, and a
superconducting transition at T$_C$ $\sim$ 1.8 K occurs. These
results are consistent with the report by Greenblatt et al. It
indicates the good quality of our single crystals. The inset
highlights the superconducting transition in the temperature range
from 0.3 K to 10 K for $Li_{0.9}Mo_6O_{17}$ single crystal along b
axis. Fig.3 (b) shows the electronic
\begin{figure}[h]
\includegraphics[width=9.5 cm]{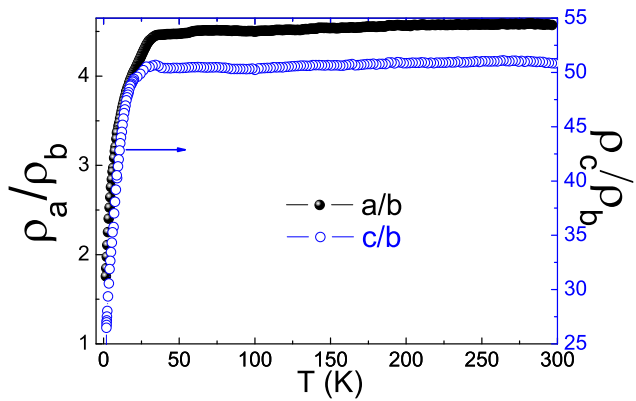}
\caption{Temperature dependence of the anisotropy
$\rho$$_c$/$\rho$$_b$ and $\rho$$_c$/$\rho$$_a$ in the temperature
range from 1.8 K to 300 K  for $Li_{0.9}Mo_6O_{17}$ single
crystal.}\label{Fig:Fig2}
\end{figure}
resistivity as a function of temperature along a-, b- and c-axis,
respectively. The values of $\rho$$_{300K}$ along a-, b- and c-axis
axis are 64.5, 16 and 854 $m\Omega cm$, respectively. The
resistivity anisotropy at 300 K $\rho$$_b$ : $\rho$$_a$ : $\rho$$_c$
$\sim$ 1 : 4 : 50. This indicates the quasi-one-dimensionality
character in $Li_{0.9}Mo_6O_{17}$. The anisotropy obtained here is
significantly different from the report by Greenblatt et al. and
Choi et al.\cite{M. Greenblatt, Choi}. However, it is close to
electrical resistivity determined by the Montgomery method.\cite{da
Luz} It should be emphasized that the anisotropy obtained here is
from the resistivity value measured on the same piece of single
crystal. Temperature dependence of the anisotropy for
$\rho$$_c$/$\rho$$_b$ and $\rho$$_c$/$\rho$$_a$ in the temperature
range from 1.8 K to 300K is shown in Fig.4. It is found that the
anisotropy of $\rho$$_c$/$\rho$$_b$ and $\rho$$_c$/$\rho$$_a$ is
temperature-independent above T$_M$. It suggests that the transport
properties' behavior of $Li_{0.9}Mo_6O_{17}$ is in someway
interconnected with all the three crystallographic directions across
the conducting chain. When electrons move between the zigzag chains
which run along the b direction through interchain hopping, it will
choose the most conducting path (for example: the
Mo6-[Mo4-O11-Mo1]-Mo6 path, where the [Mo4-O11-Mo1] path is in the
zigzag chain). It appears reasonable to conclude that the electrical
resistivities in the a- and c- axes may be $\rho_b$
dependent.\cite{da Luz} Below T$_M$, the anisotropy of
$\rho$$_c$/$\rho$$_b$ and $\rho$$_a$/$\rho$$_b$ decreases sharply.
This observation confirms $T_M$ $\sim$ 30 K as a crossover
temperature, and the sharp decrease of highly anisotropic
resistivity indicates a dimensional crossover. Through thermal
expansion experiments, Santos et al. also suggested the possibility
of a dimensional crossover at $T_M$.\cite{dos Santos} The
dimensional crossover at $T_M$ is necessary for the occurrence of
superconductivity at 1.8 K.

It is well known that classical MR in metals follows a scaling
function of (H/$\rho$)$^2$, which is known as Kohler's rule.
Kohler's rule holds if there is a single type of charge carrier and
the scattering time is isotropic on the Fermi surface. Fig.5 shows
isothermal magnetoresistance at 3 K, 5 K, 10 K, 15 K and 50 K for
the three crystallographic directions with H applied along the three
crystallographic directions, respectively. It is found that Kohler's
rule is violated in $Li_{0.9}Mo_6O_{17}$ (the curves cannot be
scaled to a universal curve). The MR behavior in
$Li_{0.9}Mo_6O_{17}$ is strongly dependent on the orientation of the
magnetic field.

\begin{figure*}[t]
\centering
\includegraphics[width=0.9\textwidth]{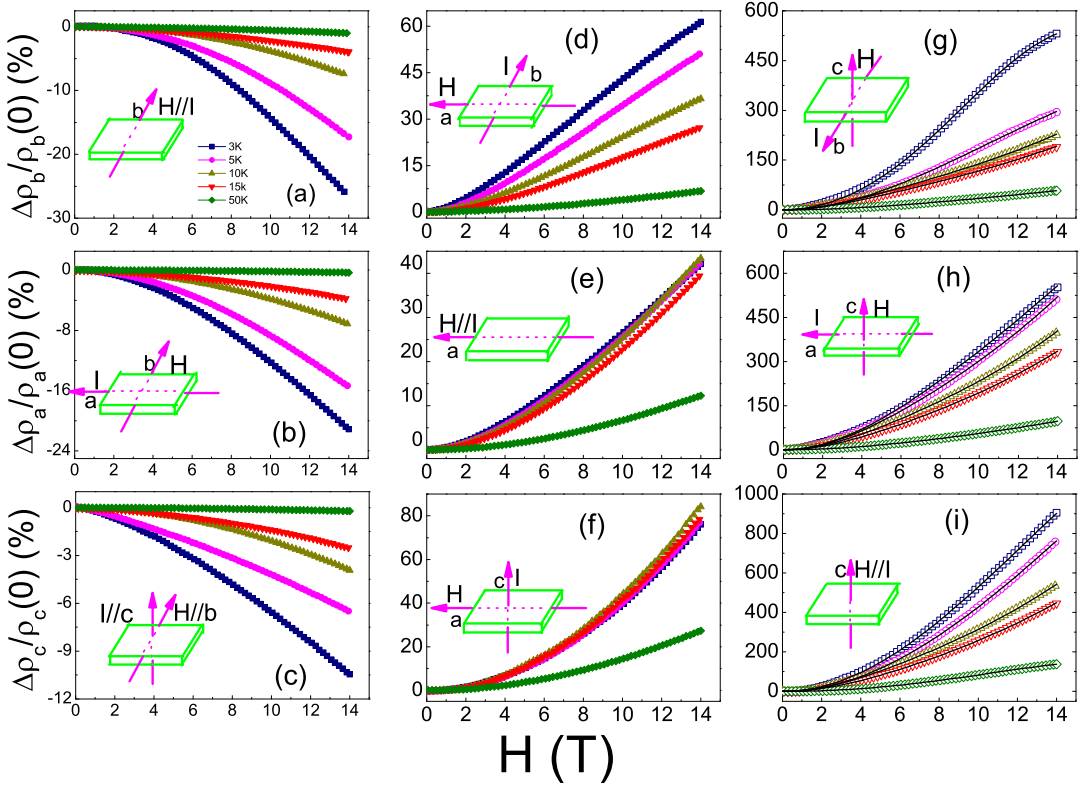}
\caption{Isothermal magnetoresistance for the three crystallographic
directions for $Li_{0.9}Mo_6O_{17}$ single crystal at 3 K, 5 K, 10
K, 15 K and 50 K, respectively, with H $\parallel$ b axis (a):
$\Delta$$\rho_b$/$\rho_b$, (b): $\Delta$$\rho_a$/$\rho_a$, (c):
$\Delta$$\rho_c$/$\rho_c$; and with H $\parallel$ a axis (d):
$\Delta$$\rho_b$/$\rho_b$, (e): $\Delta$$\rho_a$/$\rho_a$, (f)
$\Delta$$\rho_c$/$\rho_c$; and with H$\parallel$ c axis (g):
$\Delta$$\rho_b$/$\rho_b$, (h): $\Delta$$\rho_a$/$\rho_a$, (i):
$\Delta$$\rho_c$/$\rho_c$. The solid lines are the curves for the
best fit to the MR data with H $\parallel$ c axis by equation (1).}
\end{figure*}

For H $\parallel$ b-axis, the MR is negligibly small above $T_M$,
and comparatively larger negative MRs are observed below $T_M$. For
T $>$ $T_M$, the system is a pure quasi-1D, consequently, the
Lorentz force is zero with H parallel to b-axis. For T $<$ $T_M$,
the negative MR may attribute to the field suppression of an energy
gap associated with a CDW state. Recently, Xu et al. have studied
the in-chain electrical resistivity $\rho_b$(T) with field applied
along b-aixs. They considered that the negative MR is due to gap
suppression as a result of Zeeman splitting.\cite{X. Xu}

A large positive MR for the resistivity along a-, b- and c-axis is
observed with H $\parallel$ c-axis. $\Delta$$\rho$/$\rho$ = 530$\%$
$\sim$ 900$\%$ depending on different axis resistivity at 3K for H =
14 T. In CDW compounds of A$_{0.9}$Mo$_6$O$_{17}$ ( A = K, Na, Tl )
and Nb$Se_3$, large positive MR for H $\parallel$ c axis have also
been observed. It was explained by field-enhanced nesting of the
Fermi sheets, and the data can be well fitted by a modified two-band
model.\cite{Mingliang Tian, J. Q. Shen} In the model, assuming that
the two types of carrier have the same relaxation time and mass in
the two-band model of Noto and Tsuzuku,\cite{Noto} for the case of H
$\parallel$ c, the MR can be expressed as

\begin{equation}
\centering \frac{\Delta\rho}{\rho} = \frac{4(\beta-\gamma H)
\bar{\mu}^2 H^2}{(1 + \beta-\gamma H)^2 + (1 - \beta + \gamma H)^2
\bar{\mu}^2 H^2} \label{1}
\end{equation}

Here $\beta$ is the ratio of two types of carrier at H = 0.
n$_2$/n$_1$ = $\beta$ - $\gamma$ H, where $\gamma$ is a constant.
n$_1$ and n$_2$ are the concentrations of the two types of carrier.
$\bar{\mu}$ is the mobility for both types of carrier (we have
assumed that the mobilities of the two carrier types are the same).

We tried to fit the MR with H applied along c axis using equation
(1). Surprisingly, the data can be well fitted using equation (1),
as shown in figure 5 (g),(h),(i). The fitting parameters are listed
in Table 1. Therefore, the results of isothermal MR suggest the
two-band electronic structure in the $Li_{0.9}Mo_6O_{17}$ single
crystal, and it supports the existence of a CDW or SDW instability
in $Li_{0.9}Mo_6O_{17}$. Actually, a tight binding band structure
calculation for $Li_{0.9}Mo_6O_{17}$ by Whangbo et al. predicted two
fully occupied and two partially occupied Mo 4d-derived
bands.\cite{Whangbo}.

\begin{table}[h]
%\begin{minipage}[t]{2\linewidth}
\caption{ Parameters obtained by the modified two-band model to fit
the isothermal MR data with H$\parallel$ c axis as shown in Fig.5
for $Li_{0.9}Mo_6O_{17}$ at T = 3, 5, 10, 15 and 50 K. }
%\end{minipage}
\begin{center}
 \tabcolsep 0pt \vspace*{-12pt}
\def\temptablewidth{0.5\textwidth}
{\rule{\temptablewidth}{1pt}}
\begin{tabular*}{\temptablewidth}{@{\extracolsep{\fill}}ccccc}
    Temperature(K)&$\beta$ = $n_1$(0)/$n_2$(0)  & $\gamma$ & $\bar\mu$($m^2V^{-1}s^{-1}$)\\
    \hline

       3  &0.971 & 0.00027  &2.474 \\
       5  &0.965 & -0.00021 &2.197 \\
       10 &0.904 & -0.00279 &2.262 \\
       15 &0.903 & -0.00276 &1.967 \\
       50 &0.989 & 0.0047   &0.948 \\

       \end{tabular*}
       {\rule{\temptablewidth}{1pt}}
       \end{center}
        \end{table}

As the temperature decreases from 50 K, the parameter $\beta$ does
not change much, and $\gamma$ changes nonmonotonically, and
$\bar\mu$ increases sharply below $T_M$ (as listed in table 1).
$\beta$ is $\sim$ 1 at all temperature from our study, which means
$n_1$ $\approx$ $n_2$. The sharp increase of $\bar\mu$ below $T_M$
is consistent with the report by Dumas et al.\cite{J. Dumas}
Considering that n $\approx$ $\frac{2}{\rho e \bar\mu}$ ,\cite{ex}
the sharp increase of the resistivity and $\bar{\mu}$ below $T_M$
leads to the sharp decrease of  $n_1$ and $n_2$ below $T_M$. This
can be understood assuming a gap is opened at $T_M$.

The MR with H $\parallel$ a-axis is positive, much larger than the
report by Xu et al.\cite{X. Xu} The MR of b-axis (
$\Delta\rho_b/\rho_b$) monotonically increases with decreasing the
temperature, while the MR of a-axis and c-axis (
$\Delta\rho_a/\rho_a$ and $\Delta\rho_c/\rho_c$) shows different
behavior: $\Delta\rho_a/\rho_a$ and $\Delta\rho_c/\rho_c$ remarkably
increase with crossing the dimensional crossover at $T_M$, and
nearly do not change below $T_M$ and trends to saturate with
decreasing temperature just below $T_M$. Therefore, such change in
MR is closely related to the dimensional crossover. It should be
pointed out that the MR with H $\parallel$ a-axis cannot be fitted
by equation (1). As mentioned, the MR with H $\parallel$ a-axis is
larger than the report by Xu et al.\cite{X. Xu} Such difference
could arise from the disorder in $Li_{0.9}Mo_6O_{17}$ crystals. No
superconductivity is detected down to 0.6 K in Xu's report, while
superconductivity shows up at 1.8 K in our crystals. Santos et al.
found that treatment of superconducting crystals in air at 200 $^oC$
for 10 h leads to a suppression of superconductivity, indicating
that the disorder can kill the superconductivity.\cite{da Luz}
Matsuda et al. found that the superconducting transition temperature
is sensitive to the magnitude of the resistivity upturn
($R_{2K}$/$R_{min}$).\cite{Matsudat} The value of the ratio,
$R_{2K}$/$R_{min}$, is about 7 for our crystals and $T_c$ $\sim$ 1.8
K, while the value of ratio for the crystals reported by Xu et al.
is $\sim$ 20,\cite{X. Xu} so that no superconductivity is observed.
Therefore, the crystal used by Xu et al. may exist much more
disorder than the crystals studied here. Based on the above results,
it is found that a dimensional crossover occurs at T$_M$. Since
$Li_{0.9}Mo_6O_{17}$ follows Luttinger liquid at $T > T_M$, it is
likely to crossover to a CDW instability dominated by electronic
interactions below $T_M$.\cite{dos Santos,Daniel. Boies} At low
temperature, superconductivity is a competing order with the SDW
state.

\section{CONCLUSIONS}

In this paper, we have studied resistivity along the a-, b- and c-
axes in Li$_{0.9}$Mo$_6$O$_{17}$ single crystal, respectively. The
results of the anisotropy of $\rho$$_c$/$\rho$$_a$ and
$\rho$$_c$/$\rho$$_b$ confirms the quasi-one-dimensionality of the
compound. Above T$_M$ the anisotropy is temperature-independent,
while decreases sharply below T$_M$. It can be understood by a
dimensional crossover at T$_M$. For H $\parallel$ b axis, large
negative MR is observed. This may attribute to the field suppression
of energy gap associated with CDW state. Positive and large
magnetoresistance shows up with H applied along c direction. The
behavior of MR gives evidence of the existence of CDW or SDW
instability in $Li_{0.9}Mo_6O_{17}$. It indicates that the
dimensional crossover destabilizes the LL fixed point leading to a
CDW dominated by electronic interactions, as suggested by LL theory.

\vspace*{2mm} {\bf Acknowledgment:} This work is supported by the
Nature Science Foundation of China and by the Ministry of Science
and Technology of China (973 project No: 2006CB601001) and by
National Basic Research Program of China (2006CB922005)


\begin{thebibliography}{}
\bibitem{ex}
Considering $\rho$ = $\rho_1$ + $\rho_2$ = $\frac{1}{n_1e\mu_1}$ +
$\frac{1}{n_2e\mu_2}$ $\approx$ $\frac{2}{ne\bar\mu}$
$\Longleftrightarrow$ n $\approx$ $\frac{2}{\rho e \bar\mu}$, when
$\bar\mu$ and $\rho$ increase sharply below $T_M$, n will decrease
sharply.
\end{thebibliography}
\end{document}